\def\be{\begin{equation}}
\def\ee{\end{equation}}
\def\bea{\begin{eqnarray}}
\def\eea{\end{eqnarray}}
\def\bi{\begin{itemize}}
\def\ei{\end{itemize}}
\def\om{{\omega}}
\def\u{{\bf u}}
\def\a{{\bf a}}
\def\F{{\bf F}}

\def\FigRunAmp5{Fig/FigRunAmp5}
\documentclass[aps,preprint]{revtex4}
\usepackage{graphicx}

\begin{document}
\title{Using Drag to Hover}
\author{Z.~Jane Wang}
\affiliation{Theoretical and Applied Mechanics, Cornell University, Ithaca,
NY 14853}
\date{March 20, 2003}
\begin{abstract}

Unlike a helicopter, an insect can, in theory, use both lift and drag to stay aloft.
Here we show that a dragonfly uses mostly drag to hover by employing asymmetric
up and down strokes. Computations of a family of strokes further show that 
using drag can be as efficient as using lift at the low Reynolds number regime
appropriate for insects. Previously, asymmetric strokes employed by 
dragonflies were viewed as an exception. Here we suggest that these 
asymmetric strokes are building blocks of the commonly seen symmetric strokes, 
such as a figure-of-eight or a U-shape. Therefore insects which use those symmetric
strokes can also use some drag to hover. In a sense, some insects 
are rowers or swimmers in air. However unlike oars in water, insects cannot
lift their wings out of air. This leads to two subtle consequences. First, a restriction 
on the degree of asymmetry in the strokes in order to produce sufficient lift. 
It corresponding to an upper limit in the inclined angle of the stroke plane, 
about $60^0$, similar to the value observed in dragonfly flight. 
Second, a break of symmetry in the forces produced by symmetric strokes.

\verb

\end{abstract}

\maketitle

Airplanes and helicopters are airborne via aerodynamic lift, not drag.
However, it is not {\it apriori} clear that nature should design insects 
to fly using only lift. In rowing and swimming, we make use 
of drag to propel, so there is a reason to suspect that insects 
might do the same. 

It is well appreciated that micro-organisms (bacteria, sperm, and protozoa) 
use drag to  swim in low Reynolds number flows\cite{purcell77,taylor-video}.  
It is also known that some birds and fish use drag to fly and swim in 
high Reynolds number flows\cite{blake81,vogel}. But recent research on 
insect flight has primarily focused on the unsteady mechanisms for 
lift enhancement\cite{ell96,dic99}, and  seems to have overlooked 
the useful effects of drag. 

The separation of lift and drag, which are 
force components orthogonal and anti-parallel to the 
instantaneous velocity of the wing relative to the far
field flow, is natural for airplane wings, propeller 
and windmill blades, and boat sails. This is because 
large wings and sails `fly' at a relatively steady 
state and at small angles of attack, so lift is the
dominant component. But an insect uses large angles
of attack to generate high transient lift, i.e. taking
advantage of dynamic stall\cite{ell84}. 
At large angles of attack, high lift and high drag go hand in hand,
as expected for stalled flow  and as seen in recent 
experiments\cite{sane01}. Therefore the conventional 
separation of lift and drag is no longer of central interest. 
The relevant force for hovering is the net 
vertical force which balances the gravity;  thus it is 
convenient to decompose the forces  into vertical and horizontal  
components. If a wing is  restricted to move 
{\it symmetrically} along a horizontal plane, 
whether in circular helicopter-like or reciprocal motions, 
the drag roughly cancels in each cycle. And the net vertical 
force to balance an insect's weight only comes from lift.  
In contrast, {\it asymmetric} up and down strokes along an 
inclined stroke plane generate vertical forces which are 
oblique to the stroke plane. The vertical force
thus has contributions from both lift and drag.

So at least in theory, drag and lift can be  of equal use to insect.
But can using drag be as efficient as using lift at the
range of Reynolds number appropriate for insects?

Strong evidence suggesting that some insects use drag to hover
goes back to an early study by Weis-Fogh\cite{wei73} 
who noticed that true hover-flies (Syrphinae) 
and dragonflies (Odonata) employ asymmetric 
strokes along an inclined stroke plane.  This is in contrast
to the symmetric back and forth strokes near a horizontal
plane, seen in the majority of hovering insects, 
including fruit-flies, bees,  and beetles. What brought the 
asymmetric strokes to attention was the failure of quasi-steady 
calculations in predicting enough vertical forces to support an insect's
weight\cite{wei73,nor75,ell84,som85,jwang-hv}. On the other hand,
the plunging motion during the down-stroke in the asymmetric strokes
certainly indicates that large drag can be generated in the upward
direction.

Since the work of Weis-Fogh, the asymmetric strokes have been
treated separately from `normal hovering'\cite{wei73,ell84}. 
Most studies have focused on symmetric strokes along a horizontal
plane. However, commonly seen insect wing strokes can 
deviate from the horizontal plane and follow either  
parabola-like or figure-eight 
trajectories \cite{wingmotion-note}. Here we suggest that these 
strokes can be viewed, conceptually, as pairs of  asymmetric strokes,
illustrated in Fig.~1d and 1e. Because of this connection,
we can focus on asymmetric strokes  
along straight inclined planes, and deduce qualitative results for 
general symmetric strokes  used by insects.

Let us first consider two cases of hovering; one employs 
symmetric strokes along the horizontal plane, a special case of normal 
hovering, the other asymmetric strokes along an 
inclined plane of $60^\circ$, suggested by the study of dragonfly free 
flight({\it Aeschna juncea})\cite{nor75}. 
Both cases belong to the family of wing kinematics 
described by 

\bea
 [x(t), y(t)] &=& \frac{A_0}{2}(1+ \cos{2\pi f t})[\cos \beta, \sin \beta]\\
 \alpha(t) &=& \alpha_0 + B\sin(2\pi f t + \phi), \label{eq-mtn}
\eea

\noindent where $[(x(t), y(t)]$ is the position of the center of the chord,
and $\beta$ is the inclined angle of the stroke plane (see Fig.~1a-1c). 
$\alpha(t)$ describes the chord orientation relative to the 
stroke plane. $f$ is the frequency, 
$A_0$ and $B$ are amplitudes of translation and rotation, 
and $\phi$ is the phase delay between rotation and translation. 
$\alpha_0$ is the mean angle of attack, thus it describes 
the asymmetry between the up and down strokes.  
$\alpha_0=\pi/2$ and $\beta=0$ correspond to a
symmetric stroke along a horizontal plane (Fig. 1c).
The wake of this family of wing motions was visualized 
previously using smoke trajectories\cite{fre91}.

The two dimensional flow around a hovering wing is governed by
the Navier-Stokes equation, which is solved here using a fourth order 
compact finite-difference scheme\cite{eliu96}. To ensure sufficient resolution 
at the edge of the wing and efficiency in computation, 
elliptic coordinates fixed to the wing, $(\mu, \theta)$, are employed
and mapped to a Cartesian grid. The two-dimensional 
Navier-Stokes equation governing the vorticity in elliptic coordinates
is 

\begin{eqnarray}
\frac{\partial (S\om)}{\partial t} + (\sqrt{S}\u\cdot\nabla)\om
&=&\frac{1}{Re} \Delta ^2 \om \\
\nabla \cdot (\sqrt{S}\u) &=& 0,
\end{eqnarray}

\noindent where $\u$ is the velocity field, $\om$ the vorticity field, 
and $S$ the scaling factor $S(\mu,\theta)=\cosh^2\mu-\cos^2\theta$. 
The forces are calculated by integrating the fluid stress on the 
wing. Specifically, $\F=\F _p+\F _\nu + \rho W \a $, where $\F_p$ and
$\F_\nu$ are pressure and viscous contributions, given below,
$W$ is the area of ellipse, $\a$ the wing's linear acceleration,

\bea
\F _p &=&\rho\nu \int \frac{\partial \omega}{\partial \mu} 
    (\sinh\mu_0 \sin\theta{\hbox{ \^{x}}} + \cosh\mu_0 \cos\theta
    {\hbox{\^{y}}})d\theta, \\
\F _\nu &=& \rho\nu \int \omega 
(-\cosh\mu_0 \sin\theta {\hbox{ \^{x}}} + \sinh\mu_0 \cos\theta 
                      {\hbox{ \^{y}}} )d\theta.
\eea

\noindent The fictitious forces introduced by the rotating frame,
i.e.~the centrifugal and Coriolis forces, as well as the force
due to rotational acceleration, integrate to zero and
thus have no contribution. The method was described in detail
previously\cite{jwang-flp,jwang-hv}.

The instantaneous forces are nondimensionalized by $0.5\rho u^2_{rms} c$, 
where $\rho$ is the density of air, $u_{rms}$ the root mean 
square of the translational velocity of the center of the wing, 
and $c$ the chord, respectively. The dimensionless forces
are called force coefficients, $C_L$ and $C_D$ denoting the 
lift and drag coefficients,  $C_V$ and $C_H$ are
the vertical and horizontal force coefficients. Because
the horizontal force cancels over a period, its absolute
value is used when taking averages.

The translational motion of the wing is completely specified
by two dimensionless parameters, the Reynolds number,
$Re \equiv u_{max} c/\nu = \pi f A_0 c/\nu$,
and  $A_0/c$. The typical Reynolds number of a dragonfly 
is about $10^3$, and a fruit-fly is about $10^2$.  
But since the conclusions drawn below does not crucially depend on the 
Reynolds number in this range, it is chosen to be 150 for simplicity. 

In the case of  symmetric stroke (Fig.~1c and Fig. 2a), described by eq. (1) with
$\alpha_0=\pi/2$, and $\beta=0$, each half-stroke generates
almost equal lift in the vertical direction,
and almost equal drag in the opposite horizontal direction. 
The drag in each stroke does not exactly cancel even in the 
steady state due to a symmetry breaking, as will be discussed.
The averaged vertical and horizontal force 
coefficients are 1.07 and 1.61, respectively, resulting in a ratio of 0.66. 
In contrast, the asymmetric stroke
with $\alpha=60^\circ$ $\beta = 62.8^\circ$ (Fig.~3a and 3b) generates most of its
vertical force during the down-stroke, in which the lift and drag 
coefficients are 0.45 and 2.4, respectively; they are 0.50 and 0.68 
during the upstroke. The vertical and horizontal force coefficients 
averaged over one period are 0.98 and 0.75,  resulting in a ratio of 1.31,
twice the value of the symmetric stroke. In this case 76\% of the 
vertical force is contributed by  the aerodynamic drag.

Another view of the difference between the symmetric and asymmetric
strokes is revealed by the averaged flow around the wing.
The vorticity and velocity fields around the wing at four different
times are presented in Fig.~2c and 3c.  Comparing the traveling distances
of vortex pairs in the two cases suggests a faster jet produced 
by the asymmetric stroke. A better way to quantify these jets is 
to plot the time averaged velocity below the wing. The averaged flow
shows the structure of the jets, as shown in Fig.~2d and 3d. 
The velocity is plotted in physical space, which are interpretated
from the computed velocity in the body coordinates. 
The asymmetric stroke  generates a faster jet of a
width comparable to the chord, and it penetrates downward for 
about 7 chords. In contrast, the symmetric stroke generates a jet
whose width is comparable to the flapping amplitude, and 
it penetrates down for about 4-5 chords.
Thus we tentatively conjecture that dragonflies take advantage 
of the ground effects to hover above the water at a 
distance of 7 chords or less, as the jet is reflected by the water.

Next we investigate how the forces and power vary with the degree of 
asymmetry. For this  purpose, it is convenient to fix $A_0/c=2.5$, $f=1$,
$Re=150$, $B=\pi/4$,  $\phi=0$, and vary $\alpha_0$ from $\pi/4$ to $\pi/2$,
with ten  $\alpha$'s equally spaced. The time-averaged forces 
and power are plotted against $\beta$, the angle of the stroke plane, in 
Fig.~4. For a given $\alpha_0$, $\beta$ is determined such that the net force averaged 
over a period is vertically up. Fig.~4 illustrates two interesting points. 

First, as the stroke plane tilts up, the vertical force coefficient,  
$\overline{C_V}$, remains almost a constant up to $\beta \sim 60^\circ$.
The horizontal force averages to zero, but its average magnitude, 
$\overline{C_H}$ decreases with $\beta$. Thus, 
the ratio, ${\overline{C_V}}/{\overline{C_H}}$, 
increases by a factor of $2$ as $\beta$ increases from $0^0$ to $60^0$.
Therefore by employing asymmetric strokes along an inclined plane, 
an insect not only maintain the vertical force but also reduces
horizontal forces.  The averaged power exerted by the wing to the 
fluid is given by  $\overline{P}=\left<F_{D}(t) u(t)\right>$, 
where $F_{D}(t)$ is the force parallel to the translational velocity of 
the wing, $u(t)$. Comparing this power with the ideal power based on the 
actuator disk theory\cite{leishman} gives a non-dimensional measure,

\bea
  \overline{C_P}= 
\frac{\left<F_D(t) u(t)\right>}{\left<F_V(t)\right>^{3/2}}\sqrt{2\rho A_0},
\eea

\noindent  where the size of the actuator 
for a two dimensional wing is assumed to be  the amplitude $A_0$,
and $F_V$ is the vertical force. Similar 
to $\overline{C_V}$, $\overline{C_P}$ is relatively independent of $\beta$ up 
to $\beta \sim 60^\circ$. Up to $40^0$, there is a slight decrease
in power required to balance a given weight. 
Thus using drag to hover does not require extra work. 

Second, the sharp decrease in vertical forces at $\beta \sim 60^\circ$ suggests
that asymmetric strokes along nearly vertical stroke planes
would not generate sufficient force to hover. 
A quasi-steady model of the forces would not predict such a
cut-off\cite{jwang-unpublished}. The drop in vertical force
is due to the fact that an insect cannot lift its wing out of air,
so the wing must interact with its own wake. Such interaction 
reduces the net vertical force by more than half when
$\beta > 70^\circ$. Thus, one might expect that in natural flight
the angle of the  stroke plane also has an upper limit. 
Insects using inclined stroke planes appear to fall in this
category. One of the largest angles, $\beta \sim 60^\circ$,  is observed
in dragonflies ({\it Aeschna juncea})\cite{nor75};
other studies reported smaller inclined angles\cite{wak97b}.

The interaction between the wing and its wake at these Reynolds 
numbers has another consequence. Namely, symmetric 
strokes do not necessarily generate symmetric forces, as would be the case 
at zero Reynolds number flow. At the Reynolds numbers studied here,
the symmetric stroke along the horizontal plane (Fig. 1b) produces a 
slightly nonzero horizontal force. 
The symmetry is broken by the initial conditions, whose effects 
persist in time. Such persistence may seem peculiar at first, 
but can be explained as follows. The first back and forth stroke 
generates two pairs of vortices, one at the leading edge and another at 
the trailing edge. The strengths of these vortex pairs are asymmetric due 
to the  initial motion. Each newly generated vortex is paired 
with the previous vortex, thus it `remembers' the preceeding stroke. 
In other words, a sequence of vortices are 'zipped' together. 
Consequently, the  vorticity configuration at a given 
instance can be traced back to 
the first stroke, thus preserving the initial asymmetry.
The nonzero averaged horizontal force, the pwalternating magnitude 
in the peaks of the  lift (Fig. 2b),  and the bas in the average 
flow (Fig. 2c) are manifestations of this asymmetry. Similar
broken of asymmetry was observed in a previous theoretical
study of buttefly strokes\cite{iima01}

Real insect wing motions are complex and diverse, and the view of the
flapping motion taken here is simplistic. In addition, the computation
is two dimensional, but insects live in three dimensions. 
Nonetheless, the above analysis provides a new way of thinking
about the wing kinematics employed by the insects. In particular,
what was perceived as an exception in the biology literature, the
asymmetric strokes along an inclined stroke plane, can be viewed
as building blocks of more commonly seen symmetric strokes, including
the well-known figure-of-eight strokes. Moreover, the upper limit of 
the stroke plane angle predicted from the computation, $60^0$, coincides 
with the maximal angle observed in insects\cite{nor75}. It also turns out
that at least within the family of strokes studied here, using drag is as
efficient as using lift. We thus hope that biologists will investigate 
insect wing kinematics with a new question in mind, that is to what degree 
insects use drag? Our study further suggests two general lessons. 
First, to theorists, it seems to  be more natural to 
view insects as swimmers in air rather than small airplanes.
Thus instead of refining the unsteady lifting line theory for attached
flow, as was appropriate for airplanes, it is more relavent to find a better 
theory for separated flow. The classical theory of separated flow 
underpredicts the forces\cite{karman}. Second, designing  micro-scale
flapping mechanisms at very low Reynolds numbers need not follow 
the traditional rule of optimizing lift, but instead  could use both lift and drag.


{\bf Acknowledgement}
I thank S.~Childress, M.~Dickinson, C.~Ellington, and A.~Ruina, P.~Lissaman
for helpful discussions, and A.~Anderson and R. Dudley for suggestions 
on the manuscripts.  The work is supported by AFSOR, NSF, and ONR.

%

\newpage

\begin{figure}
\includegraphics[width=9cm]{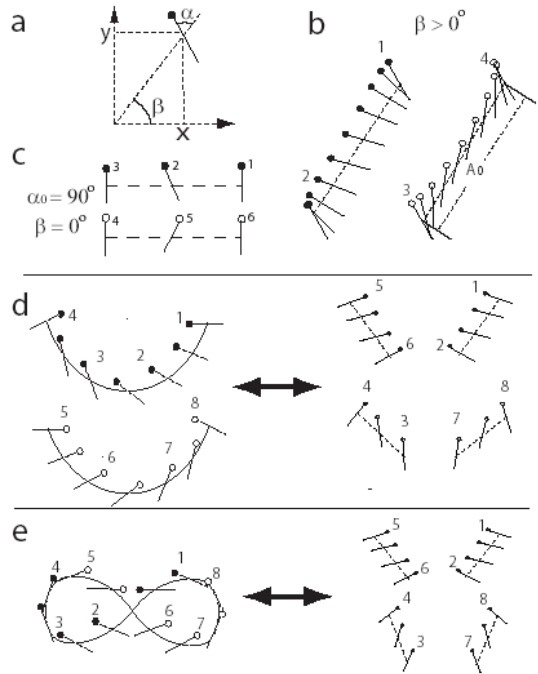}
\caption{A family of asymmetric strokes studied in this paper. Solid 
lines represent chord positions during a stroke. The leading
edge is indicated by either a black dot (down-stroke) or 
an open circle (up-stroke). The numbers next to the leading edges
indicate the time sequence. The motion is described by
eq. (1) and (2), with $A_0$ being the end to end amplitude of the stroke, 
$\beta$ the inclined angle of the stroke plane (dashed line), and 
$\alpha_0$ the initial orientation of the wing.  a) definition of 
the chord position and orientation, b) generic asymmetric strokes
c) a special case of symmetric strokes along the horizontal plane. 
d) symmetric strokes along a parabolic stroke plane, 
and e) symmetric strokes along a figure-eight.
The symmetric strokes in d) and e) can be decomposed into symmetric 
pairs of the asymmetric strokes described in b). 
The numbers next to the wings both indicate the time sequence
during a stroke, and also identifies the segments on the 
right hand side in the original stroke. This decomposition 
allows one to deduce qualitative results 
for the general symmetric strokes based on the results of 
asymmetric strokes studied here.}
\end{figure}

\begin{figure}
\includegraphics[width=12cm]{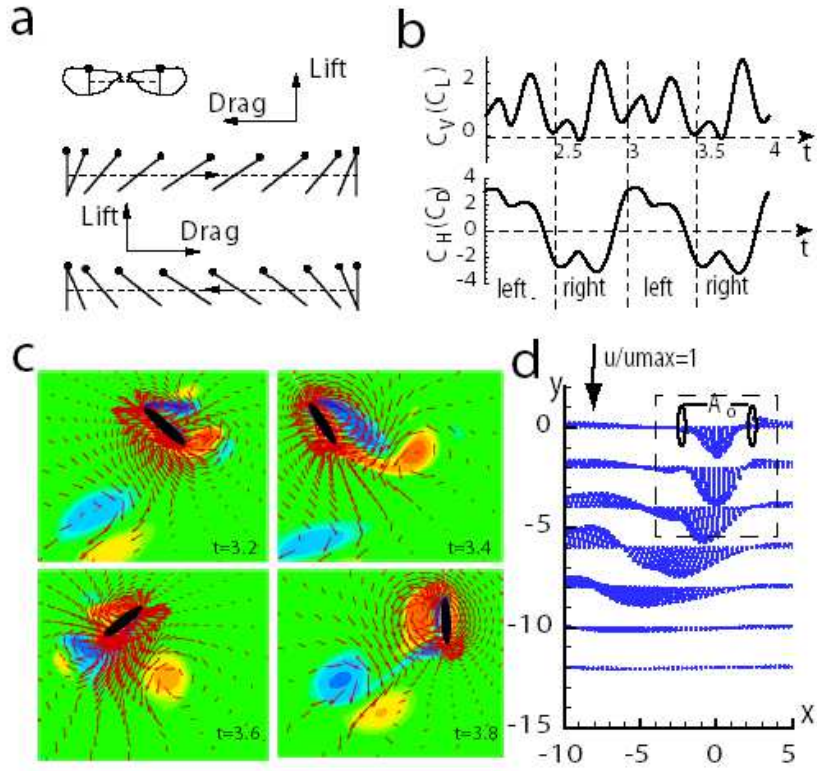}
\caption{A special case of normal hovering: symmetric strokes strictly 
along a horizontal plane. a) wing kinematics given by eq. (1) and (2)
with $A_0/c=2.5$, $\alpha_0=\pi/2$, $B=\pi/4$, $\beta=0$. The averaged 
lift and drag over each stroke are shown in scale.
In this case, drag almost cancels in two consecutive strokes, and the vertical
force is contributed primarily by aerodynamic lift. 
b) time dependent vertical ($C_V$) and horizontal ($C_H$) 
force coefficients, which are the same as lift ($C_L$) and drag 
($C_D$) coefficients.  c) snapshots of vorticity 
(red: counterclockwise rotation, blue:
clockwise rotation) and velocity field (red vectors) near the wing,
and d) time averaged velocity field over one period, characterized by
a downward jet. $u_{max}$, the maximum translational velocity
of the wing is the reference scale. The dashed square 
corresponds to the region shown in 1c).}
\end{figure}

\begin{figure}
\includegraphics[width=12cm]{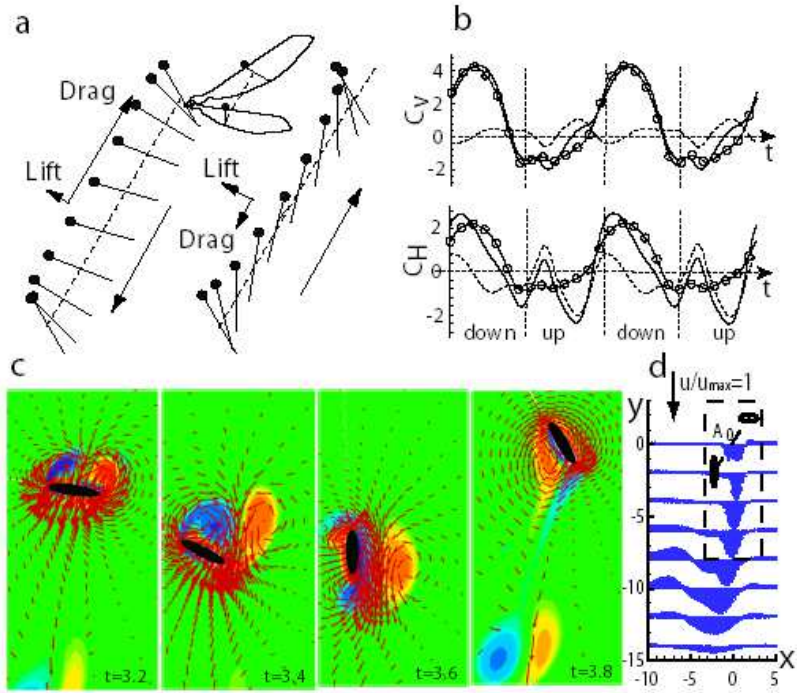}
\caption{Hovering along an inclined stroke plane using asymmetric strokes,
as seen in dragonfly flight\cite{nor75}. 
a) wing kinematics given by eq.~(1) and (2) with $A_0/c=2.5$, 
$\alpha_0=\pi/3$,  $B=\pi/4$, and $\beta=0.35\pi$. The averaged lift and 
drag over each stroke are shown in scale. In this case, the vertical 
force is contributed primarily by the drag during the down-stroke. 
b) vertical ($C_V$) and horizontal ($C_H$)  coefficients (solid lines), 
and contributions from lift (dashed line), and drag (solid line with circle), 
c)-d) see the legend of Fig 2.} 
\end{figure}

\begin{figure}
\includegraphics[width=12cm]{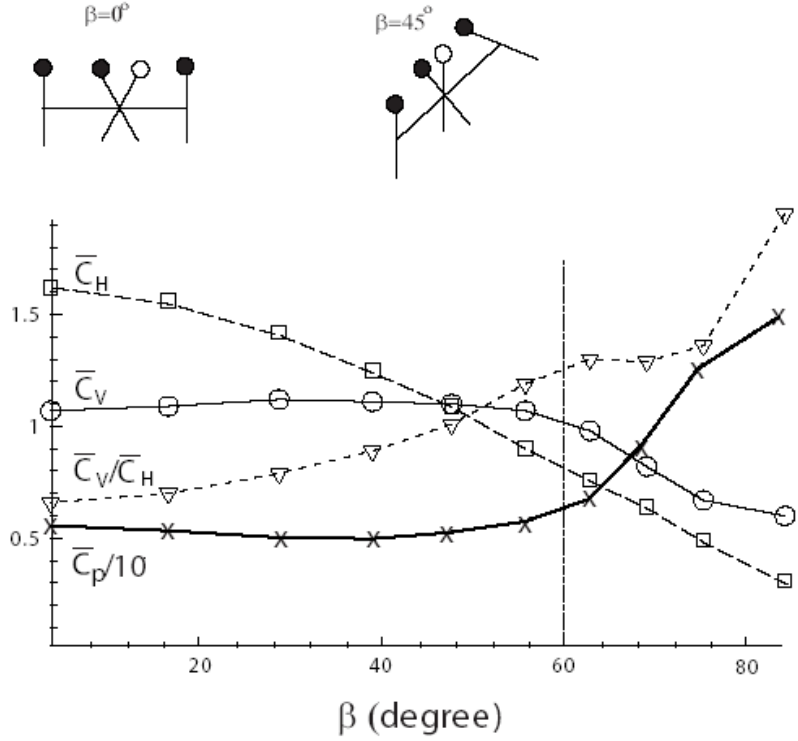}
\caption{Comparison of the averaged vertical ($\overline{C}_V$), 
horizontal ($\overline{C}_H$), and power ($\overline{C}_P$) 
coefficients as a function of the angle of the stroke plane, $\beta$,
which characterizes the degree of asymmetry in the up- and down- 
strokes and the relative contribution of lift and drag to the vertical force. 
The vertical force remains roughly constant up to $\beta=60^0$, but have
a sharp cut-off beyond that, due to the interaction between 
the wing and its own wake. The magnitude of the horizontal 
force decreases with $\beta$.  $\overline{C_P}$ is relatively
independent of $\beta$ up to $\beta \sim 60^\circ$. Up to $40^0$, 
there is a slight decrease in power required to balance a given weight. 
Thus using drag to hover does not require extra work.  }
\end{figure}

\end{document}